\documentclass{article}
\usepackage{hyperref}
\usepackage{xurl}
\usepackage{datetime}
\usepackage{graphicx}
\usepackage{natbib}
\usepackage{amsmath}
\usepackage{booktabs}

\hypersetup{colorlinks=true, allcolors=blue}

\title{A Graph Approach to the Academic Publishing Network: A Heterogeneous Model and Structural Screening over OpenAlex Open Data}

\author{Robert Šamárek \and Radek Martinek}

\newdate{articleDate}{12}{6}{2026}
\date{\displaydate{articleDate}}

\begin{document}
\maketitle
\begin{abstract}The academic publishing ecosystem forms a vast, dynamic, and heterogeneous network of interconnected entities---works, authors, institutions, journals, and topics. Traditional scientometrics, however, reduces this network to isolated tabular indicators (h-index, Impact Factor) that fail to capture its topological context and are not designed to capture coordinated illegitimate practices. Building on our companion review, which proposed a graph-based approach to publishing integrity as an unexplored direction, this paper concretizes and implements that approach. We define a heterogeneous multivariate graph model over the open data of OpenAlex (seven node types including the topical hierarchy, seven edge types), together with an analysis methodology based on projections (citation and co-authorship networks), interpretable structural metrics, community detection, and three screening detectors of anomalous publishing patterns. We deliberately avoid binary classification: the detectors return ranked candidates accompanied by explicit structural evidence for human assessment. We demonstrate the approach on a case study of the institutional corpus of VSB -- Technical University of Ostrava (2020--2025), extended by its one-hop citation neighbourhood. We show that community detection reconstructs the institution's real research groups, centralities identify cross-disciplinary bridges, and the screenings flag dense co-authorship cliques, locally closed citation loops, and thematically isolated venues. On a second, venue-centric corpus with external ground truth (journals delisted by Scopus and DOAJ) and size-matched controls, we quantitatively validate the structural features of problematic venues: we show that a naive case-control design yields seemingly strong but spurious detectors (a prominence confound), whereas after matching the only robust surviving signal is the breadth of disciplinary scope (AUC 0.70), and that an open graph-based prestige measure (PageRank over the journal citation network) tracks a JIF proxy while being an order of magnitude more resistant to citation gaming than count-based indicators. The entire analysis runs on a commodity personal computer in minutes. We release the method as the open-source library \texttt{apnet}, with a reproducible CLI workflow and an interactive web interface.\end{abstract}

\section{Introduction}

The academic publishing ecosystem is one of the largest information networks of our time: hundreds of millions of publications, tens of millions of authors, and thousands of institutions and venues form a dynamic heterogeneous graph whose structure co-determines the direction of global research \citep{fortunatoScienceScience2018, wangScienceScience2021}. Yet decisions made over this network---assessing research quality, choosing a publication venue, allocating resources---still rely largely on aggregated tabular indicators that ignore topological context: citation counts, the h-index, or the Impact Factor.

This reduction has two fundamental consequences. First, isolated indicators are manipulable---through self-citation \citep{vannoordenHundredsExtremeSelfciting2019, seeberSelfcitationsStrategicResponse2019}, citation cartels \citep{fisterTowardDiscoveryCitation2016, kojakuDetectingAnomalousCitation2021}, and the industrial production of fraudulent manuscripts by so-called paper mills, whose volume grows exponentially \citep{richardsonEntitiesEnablingScientific2025, elseFightAgainstFakepaper2021}. Second, the tabular view cannot capture phenomena that are inherently relational: cross-disciplinary bridges, the structure of research groups, or coordinated publishing patterns.

In our companion review [Šamárek \& Martinek, under review] we classified existing methods for identifying illegitimate publishing practices and showed their limitations: manual analyses and blacklists/whitelists do not scale and are inconsistent \citep{grudniewiczPredatoryJournalsNo2019}, while automated classifiers suffer from problems of accuracy, transferability, and interpretability \citep{teixeiradasilvaCanAIDetect2023}. As an unexplored direction we proposed multivariate graph analysis of the entire publishing network. The present work concretizes and implements that proposal.

A key enabler is the democratization of bibliometric data by OpenAlex \citep{priemOpenAlexFullyopenIndex2022}, which provides fully open access to the metadata of hundreds of millions of works and---as independent validation studies have confirmed \citep{culbertReferenceCoverageAnalysis2025, alperinAnalysisSuitabilityOpenAlex2024, thelwallOpenAlexSuitableResearch2025, maddiGeographicalDisciplinaryCoverage2025}---constitutes a full-fledged alternative to proprietary databases. Open data make it possible, for the first time, for any institution to build and audit an analysis of the publishing network without licensing barriers, in line with the principles of reproducible science \citep{pengReproducibleResearchComputational2011, gatesReproducibleScienceScience2023}.

This work makes three contributions:

\begin{enumerate}
\item \textbf{Model and methodology.} We define a heterogeneous multivariate graph model of the academic publishing network over OpenAlex data (seven node types, including the topical hierarchy Topic$\rightarrow$Subfield$\rightarrow$Field, and seven edge types) and an analysis methodology built on the principle of \textit{projections}: centralities and communities are computed on homogeneous projections (the citation and co-authorship networks), not on the raw mixed graph.
\item \textbf{Evidential screening, not classification.} In line with the critique of binary labels \citep{grudniewiczPredatoryJournalsNo2019, teixeiradasilvaCreditlikeRatingSystem2021}, we treat the detectors not as integrity classifiers but as reproducible structural \textit{views}: three detectors (dense co-authorship cliques, locally closed citation loops, and thematically isolated venues) rank candidates for human assessment and store concrete structural evidence with every score.
\item \textbf{A reproducible artefact.} We release the method as the open-source library \textbf{apnet} (MIT licence) with the CLI workflow \texttt{fetch} $\rightarrow$ \texttt{analyze} $\rightarrow$ \texttt{serve} and an interactive web interface; the case study on the institutional corpus of VSB-TUO is fully reproducible on a commodity personal computer.
\end{enumerate}

\section{Related work}

\subsection{Scientific networks and the science of science}

Applying network theory to science rests on solid foundations: Newman showed that scientific collaboration networks exhibit small-world properties \citep{newmanStructureScientificCollaboration2001}, and from this grew the distinct field of \textit{science of science} \citep{fortunatoScienceScience2018, zengScienceSciencePerspective2017, wangScienceScience2021}. Heterogeneous information networks became established for representing scholarly metadata \citep{shiSurveyHeterogeneousInformation2017}, with methods such as meta-path similarity \citep{sunPathSimMetaPath2011}; the direct predecessor of today's open graphs was the Microsoft Academic Graph \citep{wangMicrosoftAcademicGraph2020, farberMicrosoftAcademicKnowledge2019}, whose successor is precisely OpenAlex.

\subsection{Open bibliometric data}

Validation studies from 2024--2025 have confirmed the usability of OpenAlex for bibliometric research: reference coverage is comparable to Web of Science and Scopus \citep{culbertReferenceCoverageAnalysis2025}, OpenAlex acts as a superset of Scopus with better coverage outside Western countries \citep{alperinAnalysisSuitabilityOpenAlex2024, maddiGeographicalDisciplinaryCoverage2025}, and for research-quality assessment it provides comparable or better citation data \citep{thelwallOpenAlexSuitableResearch2025}. At the same time, data-quality limitations are documented---missing institutional affiliations in a fraction of records \citep{zhangMissingInstitutionsOpenAlex2024} and metadata incompleteness \citep{delgadoquirosCompletenessDegreePublication2024}---which must be taken into account when designing the analytics (see Discussion).

\subsection{Community detection and centralities}

For community detection, the family of modularity-based algorithms is the standard: Louvain \citep{blondelFastUnfoldingCommunities2008} and its principled successor Leiden \citep{traagFromLouvainLeiden2019}; Fortunato provides an overview \citep{fortunatoCommunityDetectionGraphs2010}. In scientometrics, community detection is used, among other things, to construct field-classification systems \citep{waltmanNewMethodologyConstructing2012}. Centralities have a long tradition in scientific networks: PageRank identifies ``scientific gems'' in citation networks \citep{chenFindingScientificGems2007} as well as influential authors \citep{dingPageRankRankingAuthors2009}, betweenness \citep{freemanSetMeasuresCentrality1977} serves as an indicator of journal interdisciplinarity \citep{leydesdorffBetweennessCentralityIndicator2007}, and k-core decomposition reveals influential cores of networks \citep{kitsakIdentificationInfluentialSpreaders2010}. We quantify interdisciplinarity in the spirit of Stirling's diversity framework \citep{stirlingGeneralFrameworkAnalysing2007, rafolsDiversityNetworkCoherence2010} and structural-holes theory \citep{burtStructuralHolesGood2004}.

\subsection{Publishing integrity and structural detection}

Structural traces of illegitimate practices are well documented: CIDRE detects citation cartels as groups of journals with anomalous mutual citation flows \citep{kojakuDetectingAnomalousCitation2021}, extreme self-citation profiles were identified by Van Noorden \citep{vannoordenHundredsExtremeSelfciting2019}, retractions associated with paper mills are rising \citep{candalpedreiraRetractedPapersOriginating2022, vannoordenMoreThan100002023}, and fabricated co-authorship networks can be identified topologically \citep{porterIdentifyingFabricatedNetworks2024}. Compared with these specialized detectors, we propose a general institutional framework in which the screening detectors are just one application of the same graph model.

\subsection{Tools}

VOSviewer \citep{vaneckSoftwareSurveyVOSviewer2010} and related tools made bibliometric mapping accessible to a broad community, but they are primarily visualization tools. Libraries such as NetworkX \citep{hagbergExploringNetworkStructure2008} or pySciSci \citep{gatesReproducibleScienceScience2023} provide an algorithmic foundation without a domain model of the publishing network. Graph databases \citep{anglesSurveyGraphDatabase2008, francisCypherEvolvingQuery2018}, in particular the embedded system Kùzu \citep{jinKuzuGraphDatabase2023}, offer a persistent, scalable layer onto which our model maps directly. \texttt{apnet} connects these layers: a domain model over OpenAlex + analytics + interactive exploration, as a single reproducible workflow.

\section{Data model}\label{sec-model}

We model the academic publishing network as a \textit{heterogeneous multivariate graph}---a directed multigraph $G = (V, E, \tau, \varepsilon)$ in which a function $\tau(v)$ assigns each node $v \in V$ a type from the set $\mathcal{T}$ and a function $\varepsilon(e)$ assigns each edge $e \in E$ a type from the set $\mathcal{R}$; both nodes and edges carry attributes (hence \textit{multivariate}). A concrete instance of the sets $\mathcal{T}$ and $\mathcal{R}$ is shown in Figure~\ref{fig-p2-schema}:

\begin{figure}[!htbp]
\centering
\includegraphics[width=0.95\linewidth]{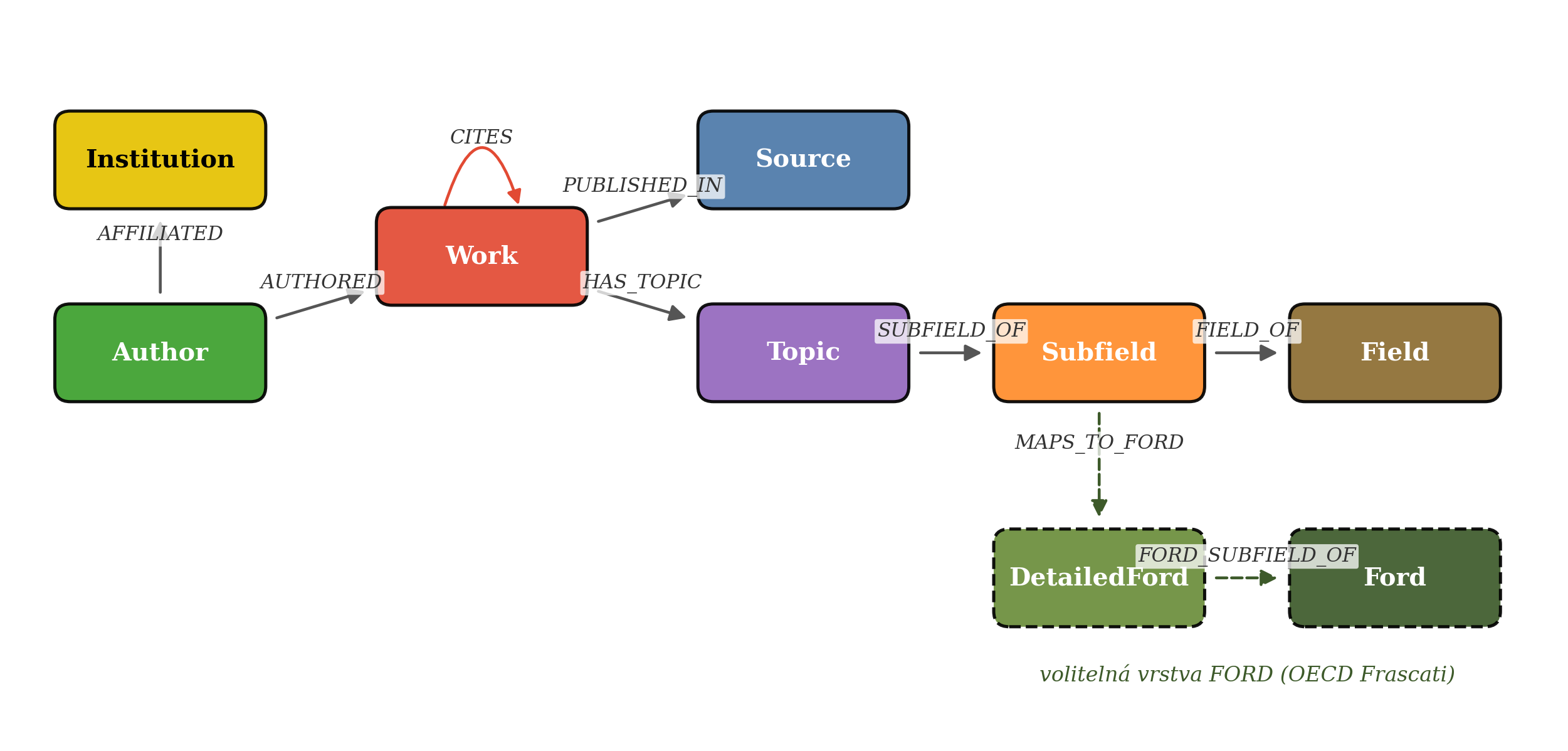}
\caption[]{Schema of the heterogeneous graph model. Seven node types (Work, Author, Institution, Source, Topic, Subfield, Field) and seven edge types. The OpenAlex topical hierarchy (Topic $\rightarrow$ Subfield $\rightarrow$ Field) is modelled as separate nodes, which enables aggregations and traversals at any level of granularity; HAS\_TOPIC edges carry a weight (the topic-assignment score). Dashed lines mark the \textbf{optional FORD layer} (OECD Frascati): DetailedFord/Ford nodes and MAPS\_TO\_FORD/FORD\_SUBFIELD\_OF edges, attached from the official OpenAlex $\times$ NORA crosswalk.}
\label{fig-p2-schema}
\end{figure}

We adopt the OpenAlex nomenclature \citep{priemOpenAlexFullyopenIndex2022}: \textbf{Work} (a publication) is the central type; \textbf{Author} is linked to works by an AUTHORED edge and to institutions by an AFFILIATED edge; \textbf{Source} (a journal, proceedings, or repository) is a work's publication venue (PUBLISHED\_IN); the topical classification forms the hierarchy \textbf{Topic} $\rightarrow$ \textbf{Subfield} $\rightarrow$ \textbf{Field} (SUBFIELD\_OF, FIELD\_OF edges), while a work may carry several weighted topics (HAS\_TOPIC with a score). Citation relationships are represented by the \textbf{CITES} edge (Work $\rightarrow$ Work). Node attributes include, among others, the publication year, the global citation count, a retraction flag (\texttt{is\_retracted}), FWCI, and open-access status; identifiers are stable OpenAlex IDs, which allows later layers (e.g., semantic representations) to be attached without remapping.

\subsection{The projection principle}

We deliberately do not run structural algorithms on the raw heterogeneous graph: paths in it mix semantically incomparable relationships (co-authorship vs. topical membership), and the resulting metrics would have no interpretation. We therefore define \textit{projections}---homogeneous graphs derived for a specific analytical question:

\begin{itemize}
\item \textbf{Citation projection} $G_C$: a directed Work $\rightarrow$ Work graph over CITES edges; it captures the flow of influence and prestige.
\item \textbf{Co-authorship projection} $G_A$: an undirected weighted Author--Author graph in which an edge weight is the number of shared works; it captures collaboration structure. Duplicate occurrences of an author on a single work are deduplicated (the projection contains no self-loops).
\item \textbf{Work $\rightarrow$ field mapping}: the transitive closure Work $\rightarrow$ Topic $\rightarrow$ Subfield $\rightarrow$ Field for topical aggregations and the computation of interdisciplinarity.
\end{itemize}

\section{Methods}\label{sec-metodika}

\subsection{Structural metrics}

We choose metrics at two levels so that each has a direct scientometric interpretation (Table~\ref{tab-metrics}).

\begin{table}
\centering
\caption[]{Structural metrics and their interpretation}
\label{tab-metrics}
\begin{tabular}{p{\dimexpr 0.333\linewidth-2\tabcolsep}p{\dimexpr 0.333\linewidth-2\tabcolsep}p{\dimexpr 0.333\linewidth-2\tabcolsep}}
\toprule
Metric & Projection & Interpretation \\
\hline
degree / strength & $G_A$ & number of collaborators / intensity of collaboration \\
PageRank & $G_C$, $G_A$ & prestige weighted by the prestige of citers \citep{chenFindingScientificGems2007, dingPageRankRankingAuthors2009} \\
k-core & $G_A$ & membership in a dense community core \citep{kitsakIdentificationInfluentialSpreaders2010} \\
weighted clustering coefficient & $G_A$ & cliquishness of an author's neighbourhood \\
betweenness (sampled) & $G_A$ & brokerage position, bridges \citep{freemanSetMeasuresCentrality1977, leydesdorffBetweennessCentralityIndicator2007} \\
interdisciplinarity (field entropy) & hierarchy & Shannon entropy of the field distribution of an author's works \citep{stirlingGeneralFrameworkAnalysing2007} \\
local vs. global citedness & $G_C$ & contrast between citations within the corpus and across all of OpenAlex \\
venue internal-citation share & $G_C$ + PUBLISHED\_IN & closedness of a journal's citation environment \\
\bottomrule
\end{tabular}
\end{table}

For betweenness we use Brandes' pivot approximation (sampling $k$ source nodes), which preserves the ranking of the most significant nodes at a fraction of the computational cost---for screening, the ranking matters more than the absolute values.

\subsection{Community detection and profiles}

We detect communities with the Louvain algorithm \citep{blondelFastUnfoldingCommunities2008} on the weighted co-authorship projection. Raw partitions are not informative in themselves; for each community we therefore construct an \textit{interpretable profile}: size, number of works in the corpus, dominant fields (via the topical hierarchy), representative authors (by number of works), and the number of retracted works. Only the profile makes it possible to confront the algorithmic partition with the reality of the institution---communities should correspond to real research groups. The architecture also allows for replacing Louvain with Leiden \citep{traagFromLouvainLeiden2019} (the interface accepts any node $\rightarrow$ community mapping); we verify the agreement of the two partitions on our corpus empirically in the results.

\subsection{Anomaly screening}

In line with the conclusions of the review, we reject the binary ``predatory/legitimate'' classification \citep{grudniewiczPredatoryJournalsNo2019}. Each detector returns \textit{ranked candidates} with a score composed of percentile ranks (robust to the heavy-tailed distributions of scientometric quantities) and with full structural evidence. The decision belongs to the human; the screening only says where to look first. We define three detectors:

\begin{enumerate}
\item \textbf{Dense co-authorship cliques} (authors): $s = \mathrm{pct}(\text{weighted clustering coef.}) \times \mathrm{pct}(\text{strength})$. Some legitimate large collaborations may exhibit a relatively star-like structure with low local clustering, whereas repeated dense cliques can be indicative of coordinated co-publication patterns---a pattern documented in fabricated networks \citep{porterIdentifyingFabricatedNetworks2024}.
\item \textbf{Locally closed citation loops} (works): $s = \mathrm{pct}\!\left(\frac{\text{citations within corpus}}{\text{global citations}+1}\right) \times \mathrm{pct}(\text{citations within corpus})$. Works intensively cited within a narrow neighbourhood but globally overlooked indicate closed citation loops \citep{fisterTowardDiscoveryCitation2016, kojakuDetectingAnomalousCitation2021}.
\item \textbf{Thematically isolated venues} (sources): $s = \mathrm{pct}(\text{internal-citation share}) \times (1 - \mathrm{pct}(\text{field entropy}))$. The combination of citation closedness and thematic narrowness is a structural trace described for captured and predatory journals.
\end{enumerate}

As a validation probe we add a \textbf{contrast of retracted works}: a comparison of the distributions of structural metrics between retracted and non-retracted works (Mann--Whitney test). This is not a classifier---retraction counts in an institutional corpus are small---but a consistency test: structural traces should be at least directionally consistent with the literature on retractions \citep{candalpedreiraRetractedPapersOriginating2022}.

\section{Implementation: the apnet library}\label{sec-implementace}

We implement the methodology as the open-source library \textbf{apnet} (Python $\geq$ 3.11, MIT licence), built on NetworkX \citep{hagbergExploringNetworkStructure2008} and pandas. The design follows three principles: (1) \textit{projections, not raw graphs}---the analytical API accepts only projections; (2) \textit{screening, not verdicts}---detectors return candidates with evidence; (3) \textit{minimal dependencies}---the core requires only NetworkX, pandas, pyarrow, requests, and scipy, while optional extras add FastAPI (interactive exploration) and Altair (charts).

\begin{figure}[!htbp]
\centering
\includegraphics[width=0.95\linewidth]{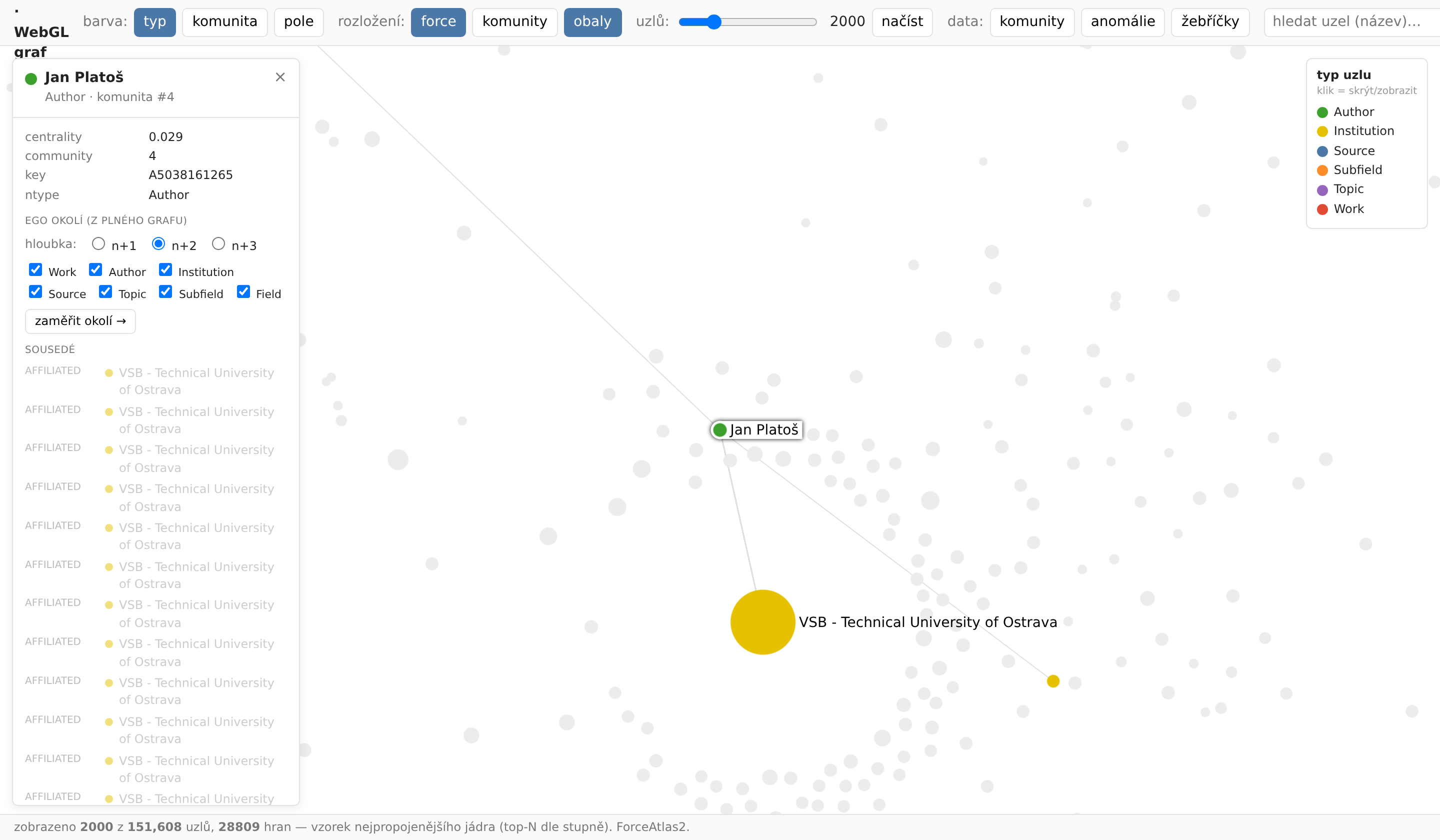}
\caption[]{The apnet explorer interactive interface (FastAPI + Sigma.js, WebGL). On the right, a force-directed (ForceAtlas2) visualization of the heterogeneous graph coloured by node type; on the left, a panel with the detail of the selected node, view switching (graph / communities / rankings / anomalies), search, and a type filter. Clicking a node shows its attributes and neighbourhood from the full graph.}
\label{fig-p2-ui}
\end{figure}

The reproducible workflow consists of three CLI commands:

\begin{verbatim}
export OPENALEX_MAILTO=you@example.org     # OpenAlex "polite pool"
apnet fetch --institution I142208455 --years 2020-2025 --out data/
apnet analyze --data data/ --out results/
apnet serve --data data/ --results results/
\end{verbatim}

The \texttt{fetch} command downloads the institution's seed corpus and its one-hop citation neighbourhood (with a configurable threshold on the minimum number of citations from the corpus, default 2---noise reduction, see Case~study:~VSB-TUO~2020–2025); data are stored as Parquet. The \texttt{analyze} command builds the graph, computes metrics, communities, and screenings, and saves the results. The \texttt{serve} command launches the web interface (Figure~\ref{fig-p2-ui}): an interactive WebGL visualization of the graph (Sigma.js) with colouring by node type or community, an ego view of a selected vertex's neighbourhood, click-through to node detail, and tabular views of rankings, communities, and screening candidates. The interface is tested headlessly (Playwright) and serves as a \textit{human-in-the-loop} tool for assessing candidates---exactly in the spirit of the screening philosophy.

\section{Case study: VSB-TUO 2020--2025}\label{sec-pripadova}

\subsection{Corpus}

The seed corpus consists of all articles affiliated with VSB -- Technical University of Ostrava (OpenAlex institution ID \texttt{I142208455}) published in 2020--2025: 7,317 works. We extend the corpus by its \textit{one-hop citation neighbourhood}: metadata of references cited at least twice from the corpus (28,094 works). The threshold $\geq$ 2 is a methodological choice that reduces noise---singly cited references form a long tail that inflates the graph without adding structural information about the repeatedly used foundations of the institution's research. The resulting graph is summarized in Table~\ref{tab-corpus}.

\begin{table}
\centering
\caption[]{The heterogeneous graph of the case study (VSB-TUO 2020--2025 + one-hop neighbourhood)}
\label{tab-corpus}
\begin{tabular}{p{\dimexpr 0.200\linewidth-2\tabcolsep}p{\dimexpr 0.200\linewidth-2\tabcolsep}p{\dimexpr 0.200\linewidth-2\tabcolsep}p{\dimexpr 0.200\linewidth-2\tabcolsep}p{\dimexpr 0.200\linewidth-2\tabcolsep}}
\toprule
Node type & Count &  & Edge type & Count \\
\hline
Work & 35,411 &  & AUTHORED & 162,187 \\
Author & 96,414 &  & AFFILIATED & 199,058 \\
Institution & 11,161 &  & CITES & 228,527 \\
Source & 5,305 &  & HAS\_TOPIC & 102,119 \\
Topic & 3,050 &  & PUBLISHED\_IN & 33,553 \\
Subfield & 241 &  & SUBFIELD\_OF & 3,050 \\
Field & 26 &  & FIELD\_OF & 241 \\
\textbf{total} & \textbf{151,608} &  & \textbf{total} & \textbf{728,735} \\
\bottomrule
\end{tabular}
\end{table}

\begin{figure}[!htbp]
\centering
\includegraphics[width=0.85\linewidth]{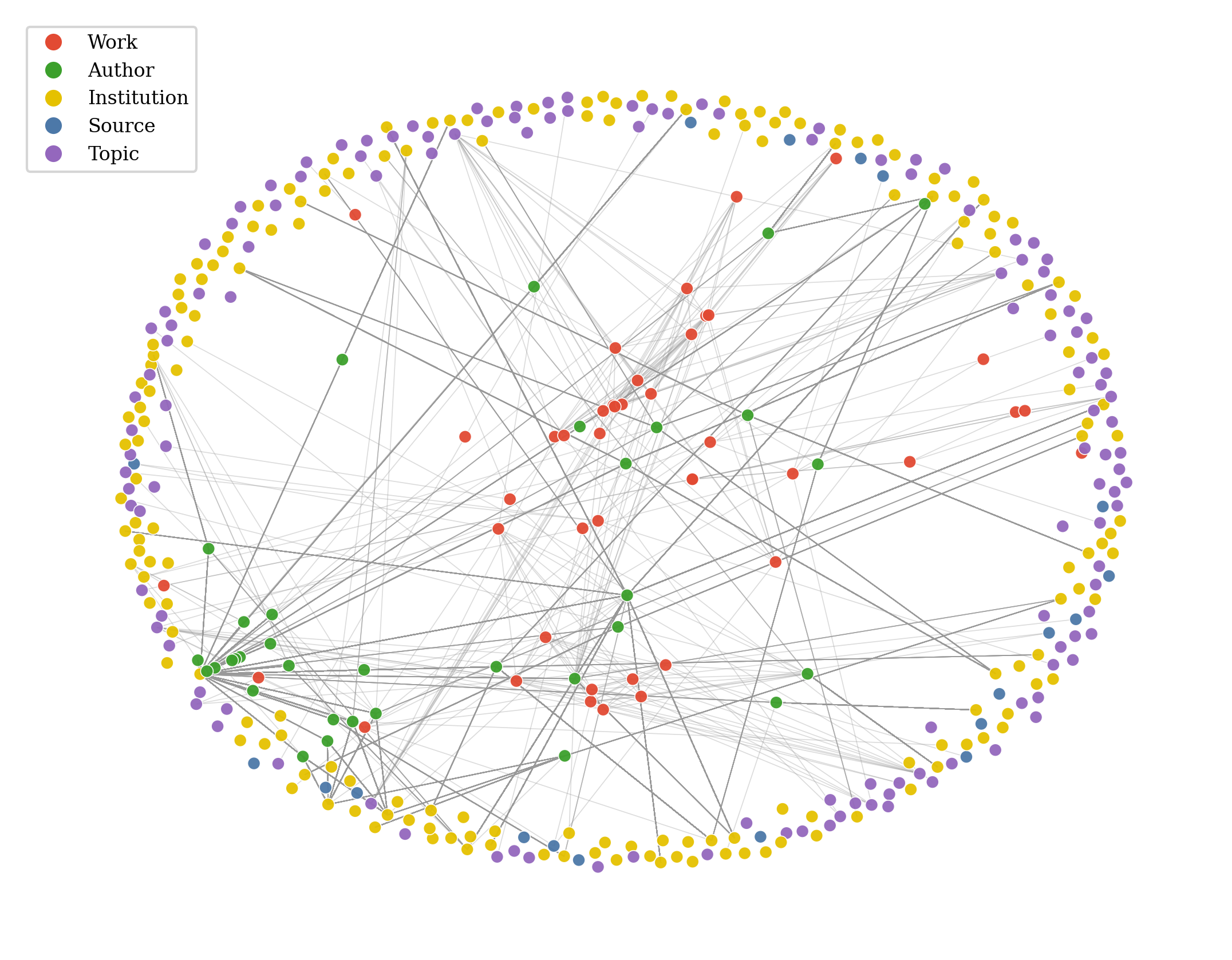}
\caption[]{A sample of the case-study heterogeneous graph (the 400 highest-degree nodes from the largest connected component). Colours correspond to node types as in Figure~\ref{fig-p2-schema}.}
\label{fig-p2-overview}
\end{figure}

The entire analysis (graph construction, metrics, communities, screenings, figures) ran on a commodity personal computer (22 cores, 30 GB RAM) in 22.8 s; downloading the corpus over the OpenAlex API takes on the order of tens of minutes and is done once. A sample of the graph structure is shown in Figure~\ref{fig-p2-overview}.

\subsection{Data quality and diagnostics}

\begin{figure}[!htbp]
\centering
\includegraphics[width=0.8\linewidth]{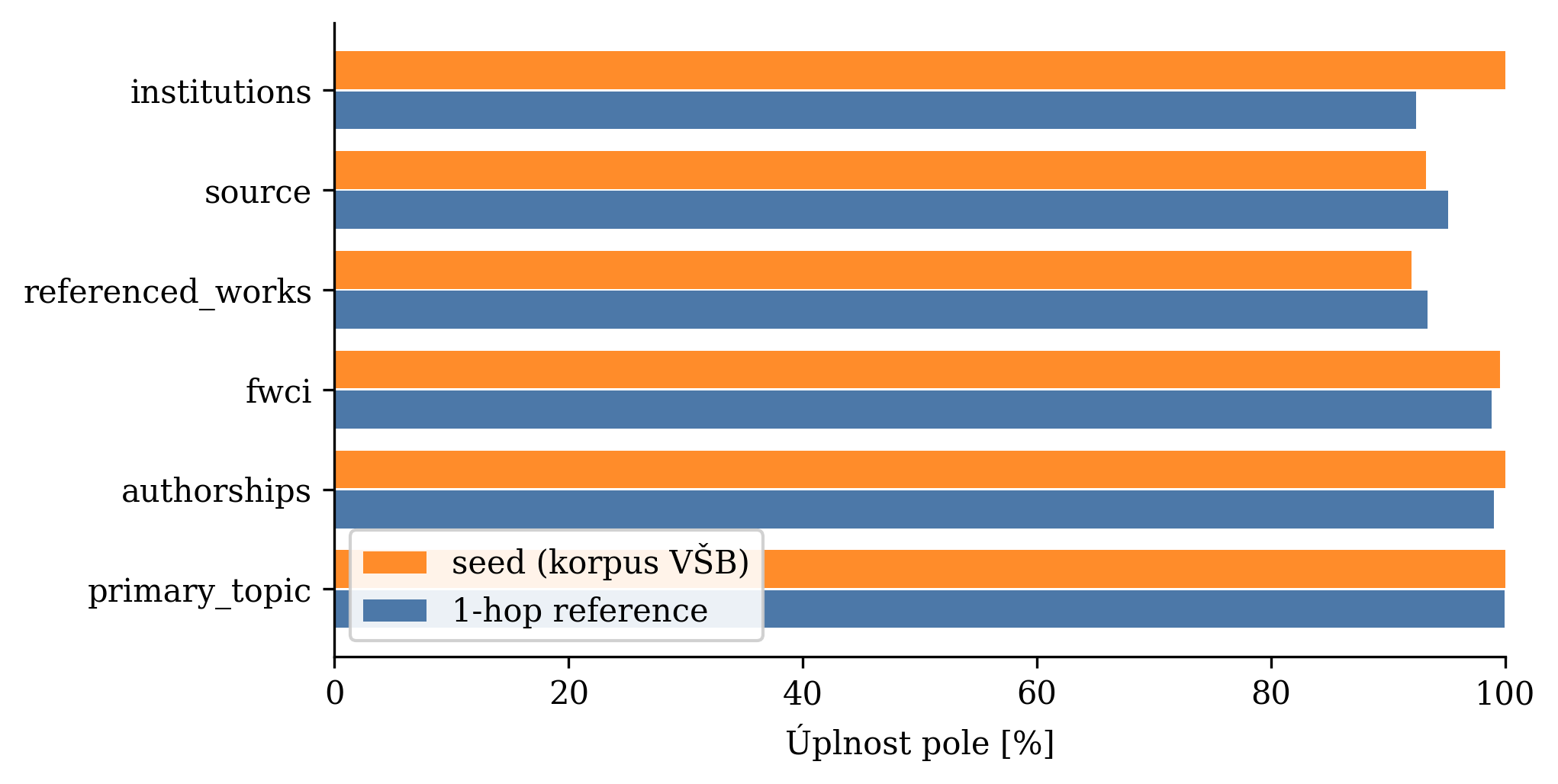}
\caption[]{Completeness of key fields in the seed corpus vs. the one-hop neighbourhood. By construction, the seed has institutions on {\textasciitilde}100\% of works (filtered on institutional lineage); the real missingness of OpenAlex shows up only in the neighbourhood.}
\label{fig-p2-completeness}
\end{figure}

Before the analysis proper, we \textbf{profile} the corpus---this is the first layer of mitigating the limitations and biases of OpenAlex \citep{zhangMissingInstitutionsOpenAlex2024, delgadoquirosCompletenessDegreePublication2024}. Profiling measures completeness, validity, consistency, and coverage and stores a machine- and human-readable report; in the apnet library it is available as a pipeline step (\texttt{apnet profile}) and enters the analysis automatically. Figure~\ref{fig-p2-completeness} summarizes the completeness of key fields for the seed corpus and its one-hop neighbourhood.

By construction, the seed corpus has institutions on {\textasciitilde}100\% of works (it is filtered by institutional lineage); the real missingness therefore appears only in the one-hop neighbourhood, where institutions are missing for 7.6\% of references---far less than the globally reported {\textasciitilde}60\% \citep{zhangMissingInstitutionsOpenAlex2024}, because the neighbourhood of a real institution consists mostly of well-indexed works. A publication venue (Source) is missing for 6.8\% of seed works. On a sample of 500 seed works, 81.6\% have an abstract---again better than the {\textasciitilde}40\% missing reported globally \citep{delgadoquirosCompletenessDegreePublication2024}. Identifiers are formally valid (0 malformed ORCIDs or ISSN-Ls). The corpus is field-concentrated (Engineering; Materials Science; Computer Science---reflecting VSB's technical profile) and geographically led by the Czech Republic (119 affiliation countries, OA share 71.6\%).

These measures do not enter the analysis merely as noise to be removed: missing metadata may correlate with the sought patterns (paper mills operate in the grey zone of metadata), and they therefore also serve as context for the screening. The risk score relies primarily on fields with high completeness; conclusions over fields with low completeness are reported with explicit uncertainty.

\subsection{Communities as research groups}

\begin{figure}[!htbp]
\centering
\includegraphics[width=0.85\linewidth]{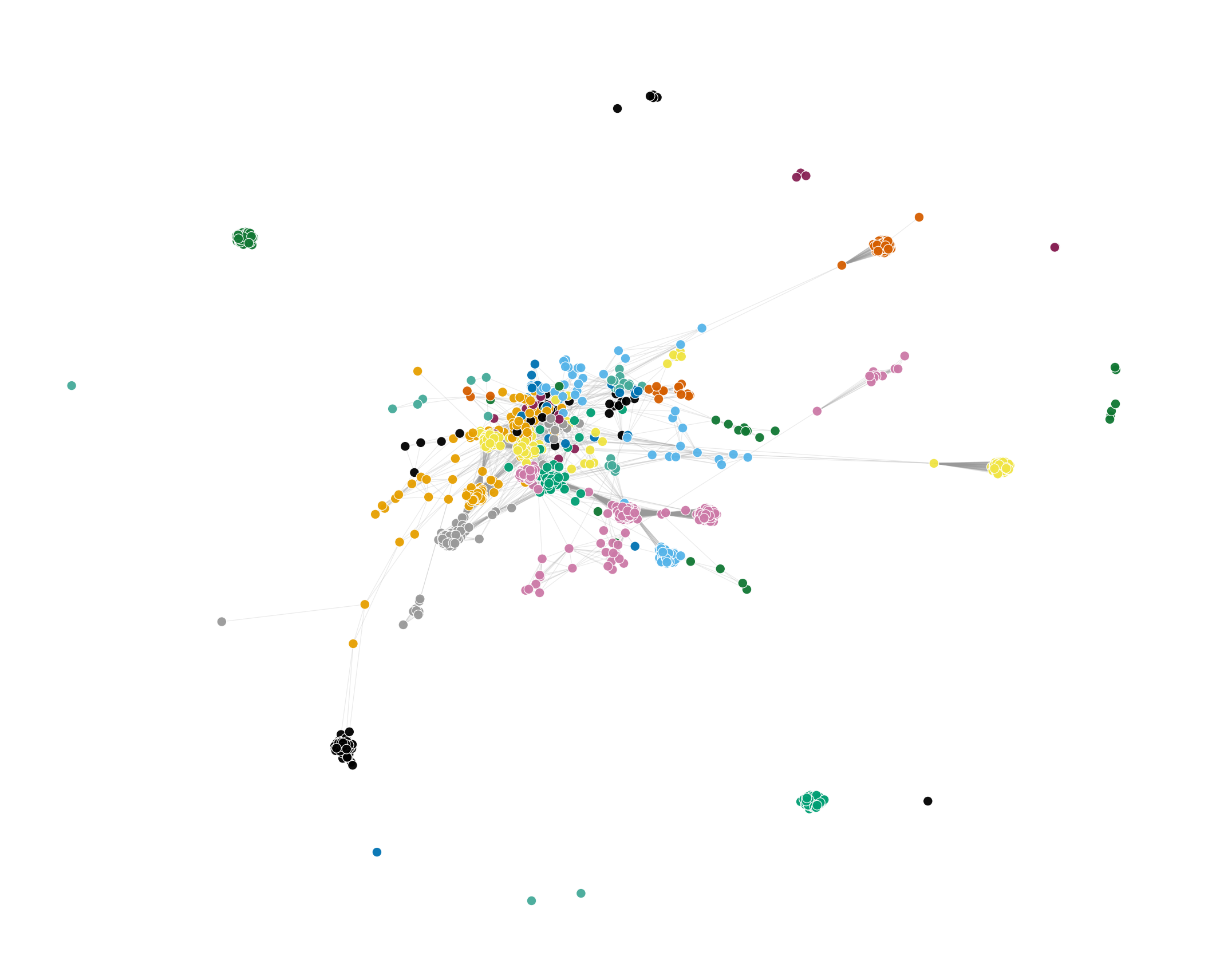}
\caption[]{The largest connected component of the co-authorship projection (49,678 authors), colour = Louvain community. The clearly separated clusters correspond to the institution's research groups (cf. Table~\ref{tab-communities}).}
\label{fig-p2-communities}
\end{figure}

The co-authorship projection has 96,414 authors and 472,306 weighted edges. Louvain detection found 11,289 communities in it (modularity 0.953); 809 communities have at least 10 members, and the largest has 3,019 authors. Figure~\ref{fig-p2-communities} shows the largest connected component coloured by community; Table~\ref{tab-communities} gives profiles of the ten largest communities.

Because Louvain can produce poorly connected communities \citep{traagFromLouvainLeiden2019}, we verify the robustness of the partition with its successor \textbf{Leiden} on the same graph: the two algorithms give practically identical partitions---the same modularity (0.953 vs. 0.953), a comparable number of communities (11,289 vs. 11,294), and high membership agreement (NMI 0.964, ARI 0.764). Conclusions about the community structure are therefore not an artefact of the algorithm choice.

\begin{table}
\centering
\caption[]{Profiles of the largest co-authorship communities (selection; full table in the repository)}
\label{tab-communities}
\begin{tabular}{p{\dimexpr 0.200\linewidth-2\tabcolsep}p{\dimexpr 0.200\linewidth-2\tabcolsep}p{\dimexpr 0.200\linewidth-2\tabcolsep}p{\dimexpr 0.200\linewidth-2\tabcolsep}p{\dimexpr 0.200\linewidth-2\tabcolsep}}
\toprule
Com. & Authors & Works & Dominant field & Representatives \\
\hline
0 & 3019 & 1075 & Engineering (472); Materials Science \dots & Kamila Kočí; Vlastimil Matějka; Dalibor M\dots \\
1 & 2676 & 922 & Engineering (618); Materials Science \dots & Róbert Čep; Jana Petrů; Radim Kocich \\
2 & 2381 & 388 & Materials Science (114); Engineering \dots & Radek Zbořil; Michal Otyepka; Aristides B\dots \\
3 & 2330 & 12 & Engineering (6); Energy (3); Material\dots & Yadong Li; Jiaguo Yu; Wenxing Chen \\
4 & 1914 & 407 & Computer Science (218); Engineering (\dots & Václav Snåšel; Seyedali Mirjalili; Laith \dots \\
5 & 1738 & 446 & Engineering (265); Computer Science (\dots & Mohit Bajaj; Vojtěch Blažek; Lukáš Prokop \\
6 & 1571 & 280 & Physics and Astronomy (99); Engineeri\dots & Muhammad Bilal Riaz; Adil Jhangeer; Nauma\dots \\
7 & 1501 & 517 & Medicine (151); Engineering (132); En\dots & Dejan Brkić; Marek Penhaker; Pavel Praks \\
\bottomrule
\end{tabular}
\end{table}

The profiles confirm the interpretability of the partition: the communities correspond to real research directions at VSB-TUO---from materials research and nanotechnology through energy to computer science---and their representative authors are the leading figures of the respective departments. This agreement is not trivial: the algorithm has no information about the organizational structure and works solely with co-authorship topology.

\subsection{Centralities: influential authors and cross-disciplinary bridges}

Table~\ref{tab-central} gives rankings of authors by PageRank (collaboration prestige) and betweenness (brokerage positions). The two views overlap only partly---and their difference is precisely what is analytically valuable: authors with high betweenness and high field entropy are the institution's \textit{cross-disciplinary bridges} \citep{leydesdorffBetweennessCentralityIndicator2007, burtStructuralHolesGood2004}, strategically important connectors that tabular indicators do not capture at all.

\begin{table}
\centering
\caption[]{The most significant authors by PageRank and betweenness (top 10)}
\label{tab-central}
\begin{tabular}{p{\dimexpr 0.143\linewidth-2\tabcolsep}p{\dimexpr 0.143\linewidth-2\tabcolsep}p{\dimexpr 0.143\linewidth-2\tabcolsep}p{\dimexpr 0.143\linewidth-2\tabcolsep}p{\dimexpr 0.143\linewidth-2\tabcolsep}p{\dimexpr 0.143\linewidth-2\tabcolsep}p{\dimexpr 0.143\linewidth-2\tabcolsep}}
\toprule
Author & Works & Degree & PageRank & Betw. & k-core & Interdisc. \\
\hline
Radek Zbořil & 212 & 656 & 0.000514 & 0.0219 & 23 & 2.4 \\
Róbert Čep & 161 & 390 & 0.000487 & 0.00753 & 13 & 1.53 \\
Muhammad Bilal Riaz & 183 & 326 & 0.000482 & 0.00777 & 8 & 2.18 \\
Václav Snåšel & 188 & 322 & 0.000457 & 0.0214 & 14 & 2.32 \\
Mohit Bajaj & 158 & 363 & 0.000449 & 0.00475 & 11 & 0.779 \\
Michal Otyepka & 206 & 528 & 0.00043 & 0.0158 & 23 & 2.45 \\
Dominik Legut & 140 & 424 & 0.000376 & 0.0141 & 30 & 1.57 \\
Martin Pumera & 304 & 269 & 0.00036 & 0.00525 & 20 & 2.39 \\
Jana Petrů & 121 & 362 & 0.000359 & 0.00784 & 14 & 1.53 \\
Radek Martínek & 193 & 320 & 0.000353 & 0.0107 & 16 & 2.34 \\
\bottomrule
\end{tabular}
\end{table}

\subsection{Anomaly screening}

\begin{figure}[!htbp]
\centering
\includegraphics[width=0.85\linewidth]{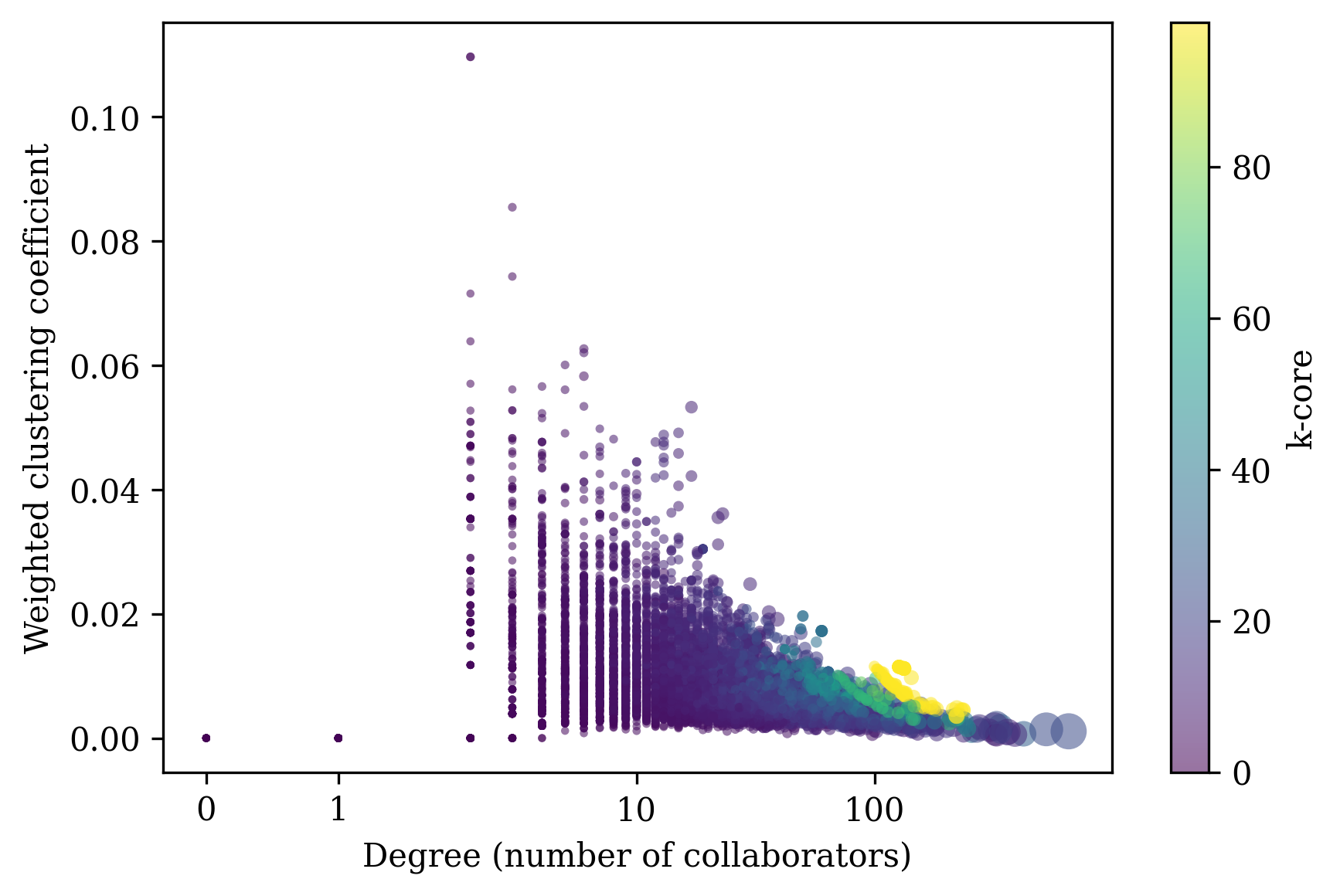}
\caption[]{Screening of dense co-authorship cliques: weighted clustering coefficient vs. degree (symlog), size = collaboration strength, colour = k-core. The typical profile decreases with increasing degree; screening candidates are the points in the top right---authors with many collaborators whose neighbourhood nonetheless remains cliquish. The interactive version in the web interface allows identification of individual points.}
\label{fig-p2-anomaly-clustering}
\end{figure}

The dense-clique detector (Figure~\ref{fig-p2-anomaly-clustering}) returns as candidates authors combining high collaboration intensity with an unusually dense neighbourhood. We emphasize the screening interpretation: dense cliques also arise legitimately (laboratory teams with stable composition); the detector only determines the ranking for human assessment, for which the web interface exposes full evidence (works, co-authors, publication venues).

\begin{figure}[!htbp]
\centering
\includegraphics[width=0.85\linewidth]{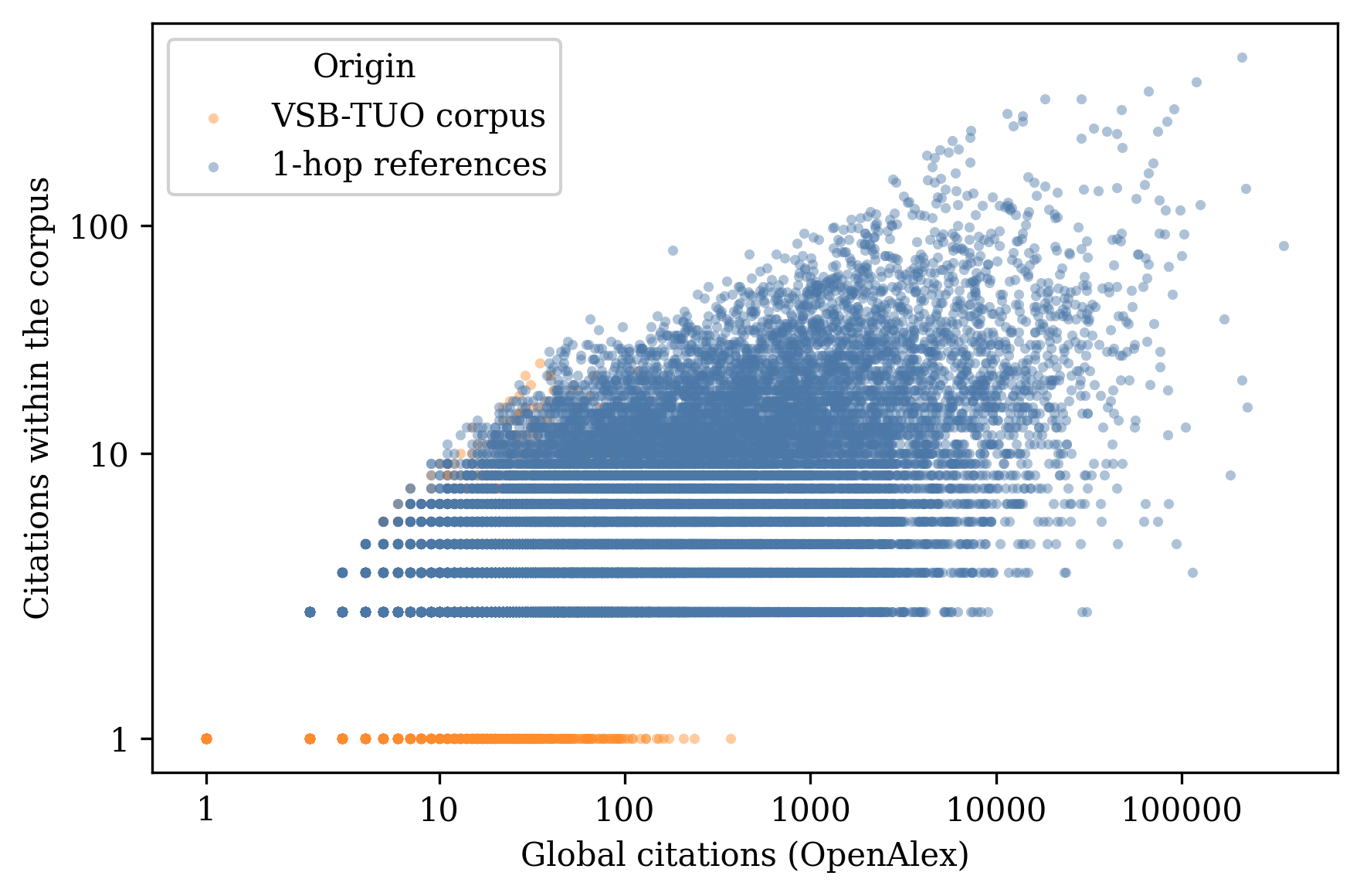}
\caption[]{Citations within the corpus vs. global citations (both axes symlog). Orange = works of the VSB-TUO corpus, blue = one-hop references. Candidates for closed citation loops lie near the diagonal---their local citedness constitutes almost all of their (modest) global interest.}
\label{fig-p2-anomaly-localglobal}
\end{figure}

The locally-closed-citation-loop detector (Figure~\ref{fig-p2-anomaly-localglobal}) contrasts citedness within the corpus with global citedness. Most works lie along the diagonal (local interest tracks global); candidates in the upper-left corner are works cited almost exclusively from their own institutional neighbourhood.

The thematically-isolated-venue detector ranked 1,668 sources; the top candidates are summarized in Table~\ref{tab-venues}. Here too the screening logic applies: a narrowly specialized legitimate journal also naturally has a high internal-citation share---the evidence (field entropy, concrete citation flows) is therefore part of the output.

\begin{table}
\centering
\caption[]{Venue screening candidates (top 10 by score)}
\label{tab-venues}
\begin{tabular}{p{\dimexpr 0.167\linewidth-2\tabcolsep}p{\dimexpr 0.167\linewidth-2\tabcolsep}p{\dimexpr 0.167\linewidth-2\tabcolsep}p{\dimexpr 0.167\linewidth-2\tabcolsep}p{\dimexpr 0.167\linewidth-2\tabcolsep}p{\dimexpr 0.167\linewidth-2\tabcolsep}}
\toprule
Publication venue & Works & Internal-cit. share & Cites out & Field entropy & Score \\
\hline
Powder Technology & 6 & 0.22 & 191 & 0 & 0.934 \\
Bulletin of the Polish Academy of Sciences Technical Sciences & 7 & 0.22 & 41 & 0 & 0.93 \\
Advances in Science and Technology -- Research Journal & 10 & 0.197 & 61 & 0 & 0.916 \\
Construction and Building Materials & 20 & 0.226 & 1273 & 0.286 & 0.912 \\
Geological Society London Special Publications & 8 & 0.185 & 466 & 0 & 0.909 \\
Journal of Magnetism and Magnetic Materials & 6 & 0.184 & 245 & 0 & 0.905 \\
International Journal of Fatigue & 6 & 0.25 & 280 & 0.65 & 0.874 \\
European Journal of Operational Research & 6 & 0.225 & 738 & 0.65 & 0.855 \\
Journal of Economic Surveys & 5 & 0.103 & 195 & 0 & 0.833 \\
Concurrency and Computation Practice and Experience & 6 & 0.176 & 34 & 0.65 & 0.822 \\
\bottomrule
\end{tabular}
\end{table}

This ranking is \textbf{not a list of problematic journals}. It only shows venues with an unusual combination of local citation closedness and thematic concentration in a given institutional corpus---it is \textit{exploratory diagnostics}, an indication of where to look first, not a verdict. That such a signal cannot be interpreted as an integrity detector without external validation is shown by the following Study C: insularity, which this screening captures, was \textbf{not confirmed} as a robust discriminating feature in a controlled design.

\subsection{Validation probe: retracted works}

The graph contains 30 retracted works (5 in the seed corpus, the rest in the one-hop neighbourhood). Table~\ref{tab-retraction} compares the medians of structural metrics. Given the small number of retractions this is a probe, not a statistical proof; the results are nonetheless directionally consistent with the literature \citep{candalpedreiraRetractedPapersOriginating2022} and illustrate how the model makes it possible to confront structural traces with external evidence.

\begin{table}
\centering
\caption[]{Contrast of structural metrics: retracted vs. other works (Mann--Whitney test)}
\label{tab-retraction}
\begin{tabular}{p{\dimexpr 0.167\linewidth-2\tabcolsep}p{\dimexpr 0.167\linewidth-2\tabcolsep}p{\dimexpr 0.167\linewidth-2\tabcolsep}p{\dimexpr 0.167\linewidth-2\tabcolsep}p{\dimexpr 0.167\linewidth-2\tabcolsep}p{\dimexpr 0.167\linewidth-2\tabcolsep}}
\toprule
Metric & Median (retracted) & Median (other) & n retr. & n other & p (MW) \\
\hline
local\_in\_degree & 2.5 & 3 & 30 & 35381 & 0.0528 \\
cited\_by\_count & 36 & 61 & 30 & 35381 & 0.236 \\
pagerank & 1.46e-05 & 1.52e-05 & 30 & 35381 & 0.338 \\
fwci & 6.35 & 4.38 & 30 & 35015 & 0.522 \\
\bottomrule
\end{tabular}
\end{table}

\section{Case study C --- controlled validation: what graph signals do and do not distinguish}\label{sec-venue}

Institutional screening shows \textit{where} to look, but does not itself say whether a pattern is problematic for integrity. The VSB case study is institutional in character, and we validate its screenings by consistency, not against ground truth. The third study deliberately addresses this gap: on a \textbf{venue-centric corpus with external ground truth} it quantitatively tests which structural features distinguish problematic venues from comparable legitimate ones---and at the same time demonstrates the \textbf{methodological trap} into which a naive case-control design falls. This study is important precisely because it refutes some intuitive signals (in particular citation insularity): it shows that the framework does not serve to confirm hypotheses, but also to falsify them.

\subsection{Corpus and ground truth}

As positives we take \textbf{delisted / discontinued venues} from two open sources: Scopus ``discontinued titles'' and DOAJ ``withdrawn'' \citep{DOAJ} (unified by ISSN); delisting from reputable indexing databases is used in the literature as a signal of problematic journals \citep{richtigPredatoryJournalsPerception2023}. For each positive we assemble \textbf{size-comparable controls} from still-indexed journals (DOAJ) so that the comparison is not distorted by output volume. For each venue we download a representative sample of works (max 200/venue), so that the within-venue citation and thematic structure is measurable at all. The resulting corpus: \textbf{341 venues} (167 positives + controls), \textbf{67,260 works} (years 2015--2023). We measure discrimination by the area under the ROC curve (AUC) with a \textbf{bootstrap 95\% confidence interval} (2,000 replications); AUC \textless  0.5 means the feature is \textit{lower} in positives than in controls.

\subsection{The prominence confound: why matched controls}

The naive variant---positives against \textbf{random} journals---yields seemingly strong detectors: median citations, FWCI, and output volume reach AUC around 0.77--0.86. Checking the direction, however, reveals a \textbf{confound}: the delisted journals were established indexed periodicals (which is why they appeared in Scopus/DOAJ), whereas the random controls contain a long tail of obscure, little-cited venues. The classifier thus detects ``was prominent,'' not ``is problematic.'' After switching to \textbf{size-matched, still-indexed} controls, these citation- and volume-based features fall to the level of chance (AUC 0.46--0.48; see Table~\ref{tab-venue-eval}). This is a separate \textbf{methodological result}: without matched controls, ``we detect predators with AUC 0.85'' would be a spurious finding reflecting prominence, not integrity.

\begin{table}
\centering
\caption[]{Discrimination of structural features between delisted and matched control venues (AUC with bootstrap 95\% CI). The ``all'' column = all positives; ``integrity'' = the subset delisted for a breach of publishing practice (DOAJ). AUC \textgreater  0.5: feature higher in positives.}
\label{tab-venue-eval}
\begin{tabular}{p{\dimexpr 0.333\linewidth-2\tabcolsep}p{\dimexpr 0.333\linewidth-2\tabcolsep}p{\dimexpr 0.333\linewidth-2\tabcolsep}}
\toprule
Feature & AUC (all) & AUC (integrity) \\
\hline
Field entropy & 0.70 [0.64; 0.75] & 0.69 [0.60; 0.79] \\
Mean FWCI & 0.45 [0.39; 0.51] & 0.66 [0.57; 0.75] \\
Median citations & 0.46 [0.40; 0.52] & 0.69 [0.59; 0.78] \\
Cites out & 0.40 [0.34; 0.47] & 0.68 [0.59; 0.76] \\
Self-citation share & 0.38 [0.32; 0.44] & 0.33 [0.24; 0.42] \\
Authors/work & 0.38 [0.32; 0.44] & 0.52 [0.42; 0.61] \\
Work count & 0.48 [0.47; 0.50] & 0.51 [0.50; 0.51] \\
\bottomrule
\end{tabular}
\end{table}

\subsection{Surviving signals}

After removing the confound, \textbf{one robust feature remains across the whole positive set}: the \textbf{breadth of disciplinary scope} (\texttt{field\_entropy}, AUC 0.70, 95\% CI [0.64; 0.75]). Delisted venues have a \textbf{broader, more dispersed} thematic scope than the matched controls (median 2.33 vs. 1.82)---a structural signature of ``scope creep.'' Conversely, insularity (self-citation share), which we originally assumed to be the main signal of predation, is definitively \textbf{dead} (AUC 0.38: positives even have \textit{fewer} self-citations).

The \textbf{integrity subset} (DOAJ ``best practice'' violations, n = 40) behaves differently: besides a broad scope, \textbf{citation impact survives} in it (FWCI AUC 0.66, median citations AUC 0.69, cites out AUC 0.68)---these venues are \textit{higher-volume and more cited} than the matched controls. This is the signature of a \textbf{``scope-creeping APC mega-journal,''} not of a low-impact predator---an empirically relevant distinction for the debate about the integrity of open-access publishing.

\subsection{Open graph-based prestige and resistance to gaming}

OpenAlex does not provide a graph-based prestige metric for journals (only the count-based \texttt{2yr\_mean\_citedness}). We reconstruct one openly: PageRank over the \textbf{citation network of venues} (in the spirit of the Eigenfactor \citep{westEigenfactorMetrics2010}, with self-cites excluded). The prestige \textbf{validly tracks} the count-based JIF proxy---Spearman 0.49 (comparable to the plain citation count, 0.51)---and is thus an open substitute for proprietary indices. Its added value is \textbf{resistance to gaming}: we inject a citation cartel (1,250 fictitious citations) into a mid-prestige venue. The count-based metric inflates \textbf{84$\times$}, whereas the graph-based prestige only \textbf{8.5$\times$} (robustness factor $\approx$ 9.9$\times$), because PageRank discounts citations from prestige-less sources. Robustness is, however, \textbf{not immunity}: a Sybil attack with many fictitious venues moves the venue from rank 172 to 12---full defence requires \textit{detecting} the cartel (a structural anomaly), which connects the metric substitute with the integrity line.

\subsection{Limitations of Study C}

The evaluation is in-sample, and the matching is so far only by size and indexing status (not by field or age); the integrity subset is small (n = 40, wide CIs). A multivariate score with cross-validation and field matching are future work. The venue citation network is sparse (only within-corpus edges), so the correlation of prestige with JIF is rather a lower bound. Nevertheless, the study provides what the VSB study lacks: a \textbf{quantitative, ground-truth-based validation} with sober effects (AUC 0.66--0.69) and an explicit methodological warning against the prominence confound.

\section{Discussion}

\textbf{What the approach delivers.} The case study documents three qualitative shifts relative to tabular scientometrics. First, \textit{structural interpretability}: community profiles reconstruct real research groups without knowledge of the organizational structure, thus giving the institution an independent picture of its own collaboration topology. Second, \textit{relational indicators}: cross-disciplinary bridges (betweenness $\times$ field entropy) are, in principle, unavailable to indicators computed per author. Third, \textit{transparent screening}: anomaly candidates are backed by concrete structural evidence that can be explored directly in the web interface---unlike black-box classifiers \citep{teixeiradasilvaCanAIDetect2023}, every score is decomposable into interpretable components.

\textbf{Limitations.} (1) \textit{Data quality}: OpenAlex exhibits missing affiliations \citep{zhangMissingInstitutionsOpenAlex2024} and incomplete metadata \citep{delgadoquirosCompletenessDegreePublication2024}; the screening scores therefore work with percentile ranks, which are more robust to missing values, but do not provide full immunity. (2) \textit{Corpus delimitation}: a one-hop neighbourhood with a threshold $\geq$ 2 is a pragmatic choice; citation loops operating just beyond the corpus boundary may escape. (3) \textit{Ground truth}: in the institutional study we validate screenings by consistency (communities vs. real groups, the retraction probe), not by a classification metric---which is in line with the review's conclusion that reliable binary ground truth for ``predation'' does not exist; Study C shows that even where external ground truth is available (delisted journals), it leads to defensible conclusions only in combination with matched controls, otherwise it measures prominence instead of integrity. (4) \textit{Scale}: NetworkX suffices for institutional corpora; for national or global analyses, a mapping of the model onto the embedded graph database Kùzu \citep{jinKuzuGraphDatabase2023} and parallel implementations of the algorithms are prepared.

\textbf{Relation to follow-up work.} The model is designed as the structural layer of a broader framework: stable OpenAlex identifiers make it possible to attach a semantic layer (vector representations of documents) and combine structural and content similarity in hybrid queries. We develop this fusion in the follow-up paper.

\section{Conclusion}

We built on the review article, which proposed graph analysis of the publishing network as an answer to the limitations of both manual lists and black-box classifiers, and presented a concrete realization of it: a heterogeneous multivariate model over the open data of OpenAlex, a methodology of projections and interpretable metrics, community profiles, and three screening detectors of anomalous patterns---all as the open-source library apnet with a reproducible workflow and interactive interface.

The VSB-TUO case study showed that the approach is feasible on commodity hardware and that its outputs are interpretable and can be confronted with reality: communities correspond to research groups, centralities identify bridges, and screenings direct attention in a structured way. The institution thus gains a research-intelligence tool that complements (rather than replaces) expert judgement. The venue-centric Study C complemented this picture with a quantitative, ground-truth-based validation and a methodological warning: without size-matched controls, a naive case-control design measures prominence instead of integrity; after matching, the breadth of disciplinary scope remains a robust feature of problematic venues, and open graph-based prestige proves to be an order of magnitude more resistant to citation gaming than count-based indices.

Future work will proceed in three directions: (1) scaling to national and global corpora (Kùzu, Leiden, parallel centralities), (2) extending the screenings with specialized cartel detectors of the CIDRE type \citep{kojakuDetectingAnomalousCitation2021} and temporal analysis, and (3) integrating a semantic layer for hybrid structural--content queries.

\section{Appendix: case study B --- a worldwide LLM corpus}\label{sec-appendix-llm}

The second case study verifies the \textbf{transferability} of the approach: the same apnet library and the same analysis script, but an entirely different---\textbf{thematically delimited and worldwide}---corpus, without any change to the code. The seed consists of works with the phrase ``large language model'' in the title, published in the last 6 months (type article): 12,520 works; the one-hop neighbourhood is restricted to the \textbf{canonical foundations} (references cited $\geq$ 10$\times$ from the corpus, 683 works---works such as ``Attention Is All You Need,'' BERT, GPT). The resulting graph has 80,055 nodes and 183,421 edges.

\textbf{A different data-quality regime.} Profiling reveals a picture markedly different from the institutional corpus: institutions are present on only \textbf{59.4\%} of seed works (versus {\textasciitilde}100\% in the VSB corpus)---i.e., they are missing for \textbf{40.6\%} of works; this is substantial missingness, though \textbf{lower} than the globally reported {\textasciitilde}60\% \citep{zhangMissingInstitutionsOpenAlex2024}. The same platform thus exhibits very different completeness depending on the type of selection. Abstracts are present on 83.2\% of works (sample), the OA share is 74.0\%, and affiliations cover 137 countries with a dominance of CN (13,497) and US (10,358). Diagnostics are therefore not a one-off calibration but a \textbf{per-corpus} step.

\begin{figure}[!htbp]
\centering
\includegraphics[width=0.85\linewidth]{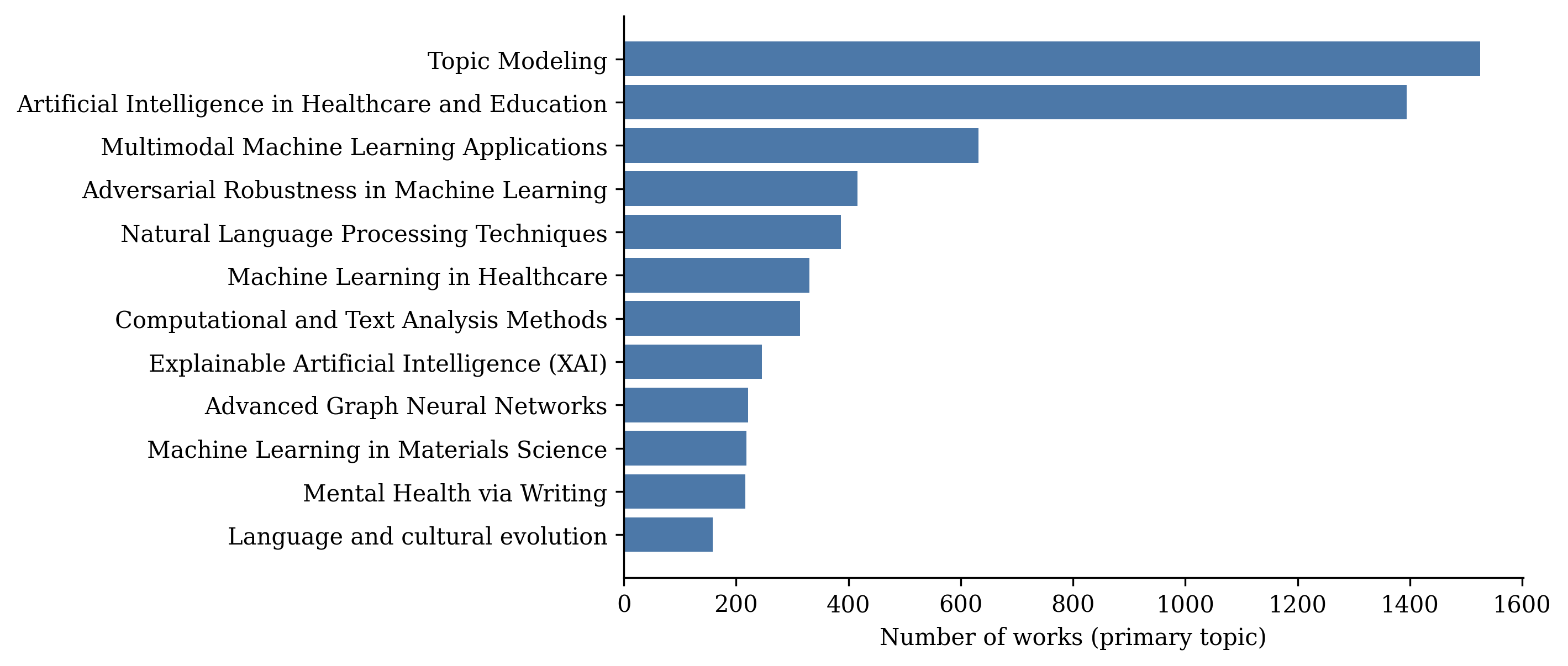}
\caption[]{The most frequent primary topics (OpenAlex Topic) in the worldwide corpus of works on LLMs over the last 6 months---a factual picture of what LLM research currently addresses.}
\label{fig-llm-themes}
\end{figure}

\textbf{A different structural regime.} The co-authorship network is strongly \textbf{fragmented}---the largest connected component contains only 5,925 of 55,677 authors (10.6\%), in contrast to the dense core of the institutional corpus; modularity 0.987. The citation signal is moreover \textbf{immature}: only 2.5\% of seed works are cited by another work within the corpus (recent works have not yet accumulated citations). This is instructive in itself---for fresh thematic corpora, structural information is carried more by co-authorship and thematic composition than by citations, and the canonical backbone must be supplemented with the one-hop neighbourhood. The thematic composition of LLM research is summarized in Figure~\ref{fig-llm-themes} (dominated by Topic Modeling; Artificial Intelligence in Healthcare and Education; Multimodal Machine Learning Applications; Adversarial Robustness in Machine Learning).

\begin{figure}[!htbp]
\centering
\includegraphics[width=0.8\linewidth]{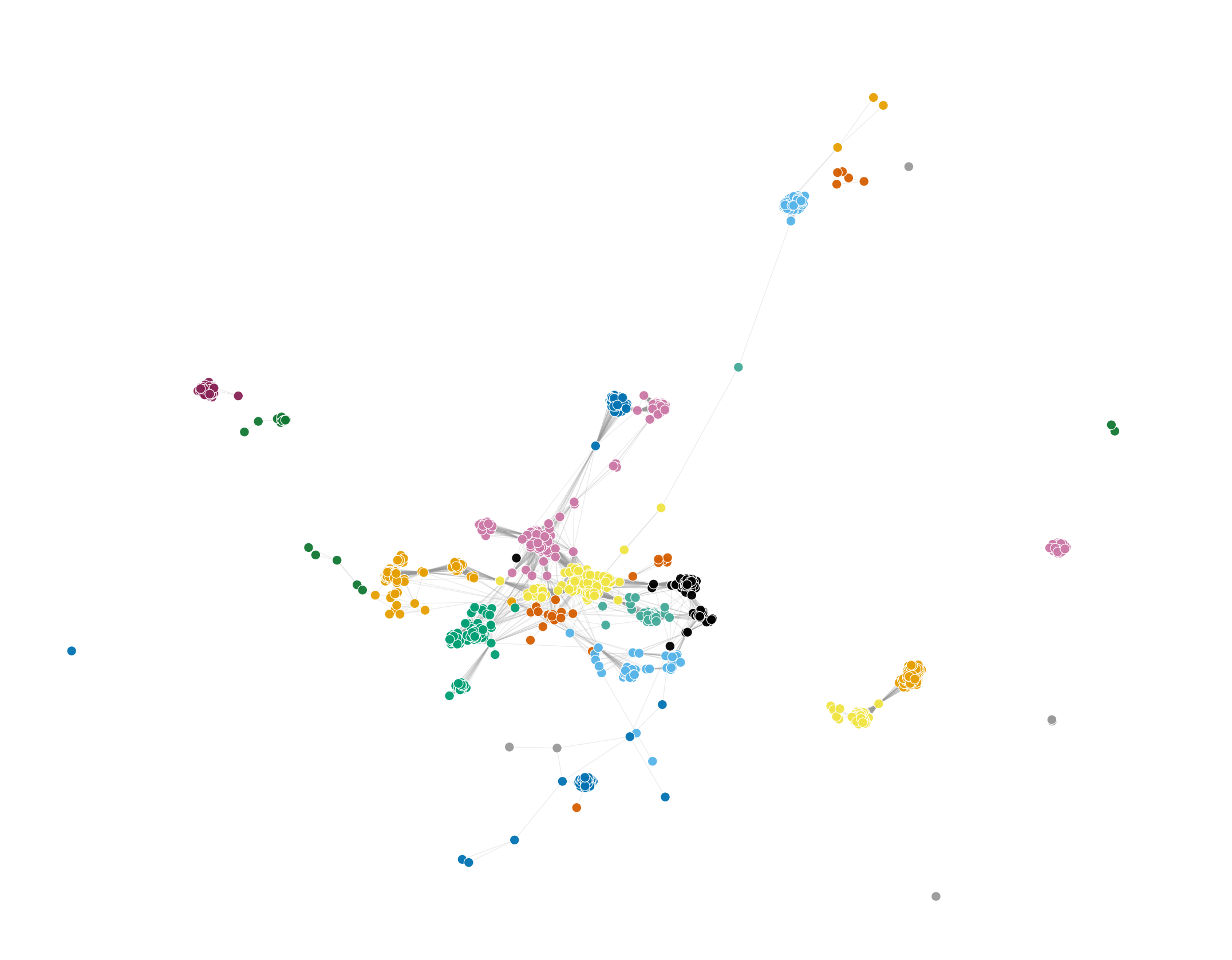}
\caption[]{The largest connected component of the co-authorship network of the LLM corpus (sample), colour = community. Compared with the institutional corpus, the network is markedly more fragmented.}
\label{fig-llm-communities}
\end{figure}

The appendix's conclusion: the same framework, without modification, transferred the analysis from a single institution to a worldwide thematic corpus, revealed \textbf{different regimes} of data quality and topology, and confirmed the necessity of per-corpus diagnostics. This supports the claim of a reproducible, generally applicable artefact.

\section{Data and code availability}

The apnet library is available under the MIT licence, including tests, documentation, and the web interface. All case-study data come from the public OpenAlex API; the repository contains the exact commands to re-download them (\texttt{apnet fetch}) as well as the complete analysis script that generates all the figures, tables, and numerical values of this article from a single source. The analysis requires no credentials (the OpenAlex API is open; the recommended contact e-mail serves only the polite pool).

\bibliography{main}
\end{document}